\documentclass[conference]{IEEEtran}
\pdfoutput=1

\usepackage{amsmath}
\usepackage{amsthm}
\usepackage{amssymb}
\usepackage{bm}
\usepackage{xspace}
\usepackage{xcolor}
\usepackage{graphicx}
\usepackage{url}
\usepackage{framed}
\usepackage{float}
\usepackage{rotating}
\usepackage{verbatim}
\usepackage{listings}
\usepackage{lscape}

\usepackage{pgfplots}
\usepackage{pgf}
\usepackage{tikz}
\usetikzlibrary{arrows,shapes.misc,chains,scopes}
\usepackage{pgfplotstable}
\usepackage{colortbl}
\usetikzlibrary{matrix}
\usetikzlibrary{calc}
\usetikzlibrary{fit}

\newcommand{\executeiffilenewer}[3]{%
\ifnum\pdfstrcmp{\pdffilemoddate{#1}}%
{\pdffilemoddate{#2}}>0%
{\immediate\write18{#3}}\fi%
}

\graphicspath{{figures/}}

\newcommand{\vecb}{\boldsymbol{b}}

\newcommand{\vecc}{\boldsymbol{c}}

\newcommand{\vecu}{\boldsymbol{u}}

\newcommand{\setr}{\ensuremath{\mathbf{R}}\xspace}

\newcommand{\setc}{\ensuremath{\mathcal{C}}\xspace}

\newcommand{\setx}{\ensuremath{\mathcal{X}}\xspace}

\newcommand{\bmm}{\begin{matrix}}
\newcommand{\emm}{\end{matrix}}
\newcommand{\bpm}{\begin{pmatrix}}
\newcommand{\epm}{\end{pmatrix}}

\newcommand{\bsbm}{\left[\begin{smallmatrix}}
\newcommand{\esbm}{\end{smallmatrix}\right]}

\newcommand{\bbm}{\begin{bmatrix}}
\newcommand{\ebm}{\end{bmatrix}}

\DeclareMathOperator*{\argmax}{argmax}

\DeclareMathOperator{\expop}{E}
\DeclareMathOperator{\entop}{H}
\DeclareMathOperator{\miop}{I}

\usepackage{booktabs}
\usepackage{multirow}
\usepackage{siunitx}
\usepackage{cite}

\usetikzlibrary{shapes}

\newcommand{\rbiawgn}{\mathsf{R}_\textnormal{biAWGN}}

\newcommand{\rate}{\mathsf{R}}

\DeclareMathOperator{\atanh}{atanh}

\newcommand{\rlm}{\rate_\textnormal{LM}}

\newcommand{\nc}{n}%{n_{\!\mathsf{c}}}

\newcommand{\qbi}{\mathsf{q}}

\newcommand{\logn}{\log_2\!\nc}

\newtheorem{lemma}{Lemma}
\newtheorem{remark}{Remark}

\newcommand{\tikzsetnextfilename}[1]{}

\title{Efficient Polar Code Construction\\for Higher-Order Modulation}

\author{Georg~B\"ocherer, Tobias Prinz, Peihong Yuan, Fabian Steiner\\Institute for Communications Engineering\\Technical University of Munich\\
Email: \{georg.boecherer,tobias.prinz,peihong.yuan,fabian.steiner\}@tum.de}

\begin{document}

\maketitle

\begin{abstract}
An efficient algorithm for the construction of polar codes for higher-order modulation is presented based on information-theoretic principles. The bit reliabilities after successive demapping are estimated using the LM-rate, an achievable rate for mismatched decoding. The successive demapper bit channels are then replaced by binary input Additive White Gaussian Noise (biAWGN) surrogate channels and polar codes are constructed using the Gaussian approximation (GA). This LM-rate Demapper GA (LM-DGA) construction is used to construct polar codes for several demapping strategies proposed in literature. For all considered demappers, the LM-DGA constructed polar codes have the same performance as polar codes constructed by Monte Carlo (MC) simulation. The proposed LM-DGA construction is much faster than the MC construction. For 64-QAM, spectral efficiency 3 bits/s/Hz, and block length 1536~bits, simulation results show that LM-DGA constructed polar codes with cyclic redundancy check and successive cancellation list decoding are \SI{1}{dB} more power efficient than state-of-the-art AR4JA low-density parity-check codes.
\end{abstract}

%%%%%%%%%%%%%%%%%%%
%%% INTRODUCTION
%%%%%%%%%%%%%%%%%%%

\section{Introduction}

Polar codes were proposed in \cite{stolte2002rekursive,arikan2009channel} and it was proven in \cite{arikan2009channel} that they achieve the capacity of binary input discrete memoryless channels, asymptotically in the block length. The principle of polar codes is as follows. The channel input bits are transformed by a polar transformation into polarized bits. Under successive cancellation (SC) decoding, the polarized bits are either very reliable or very unreliable. The unreliable bits are frozen to predetermined values and the reliable bits are used for information transmission. The frozen bit positions define a polar code.

Polar code construction consists in choosing the frozen bit positions. Monte Carlo (MC) construction estimates the quality of the bit channels by extensive simulation \cite[Sec.~6.1]{stolte2002rekursive},\cite{arikan2009channel} and is computationally demanding. MC construction for the binary input additive white Gaussian noise channel (biAWGN) is discussed in \cite{vangala2015comparative}. In \cite{mori2009performance}, polar codes are constructed using density evolution. Approximate density evolution based on Gaussian approximations (GA) is considered in \cite{trifonov2012efficient,vangala2015comparative,dosio2016polar}. The authors of \cite{trifonov2012efficient,vangala2015comparative} use the $\phi(x)$ function introduced in \cite{chung2001analysis} while \cite{dosio2016polar} uses the $J$-function \cite{tenbrink2004design} with the approximation \cite[Eqs.~(9), (10)]{brannstrom2005convergence}.

In polar-coded modulation (PCM) \cite{seidl2013polar} for constellations with $2^m$ signal points, a successive demapper (`polar demapper') connects $m$ binary polar codes to the $m$ bit levels of the channel inputs, see Fig.~\ref{fig:pcm}. MC construction for PCM was considered in \cite{mahdavifar2016polar}. The conventional GA construction can be applied to PCM by replacing the bit levels of the channel inputs by biAWGN surrogate channels (for a review of code design via surrogate channels, see \cite[Sec.~IV]{steiner2016protograph} and references therein). We call this approach the Channel GA (CGA) construction. The authors of \cite{seidl2013polar} propose to characterize the bit channels of the polar demapper by mutual information (MI). They then replace the polar demapper bit channels by biAWGN surrogate channels and use GA construction. We call this method the MI demapper GA (MI-DGA) construction. The MI-DGA construction was recently used in \cite{tavildar2016bit}. Several polar demappers proposed in literature (e.g., in the context of bit-interleaved coded modulation (BICM)), first calculate bit-wise log-likelihood ratios (LLR), which are then processed as independent. However, LLRs calculated from the same channel output are dependent. Such demappers are therefore \emph{mismatched} (`MM') and as we will show in this work, the MI-DGA construction does not work well for mismatched demappers.
\begin{figure}
\footnotesize
\centering
%\tikzsetnextfilename{mainplot}
%\input{figures/mainplot}
\includegraphics{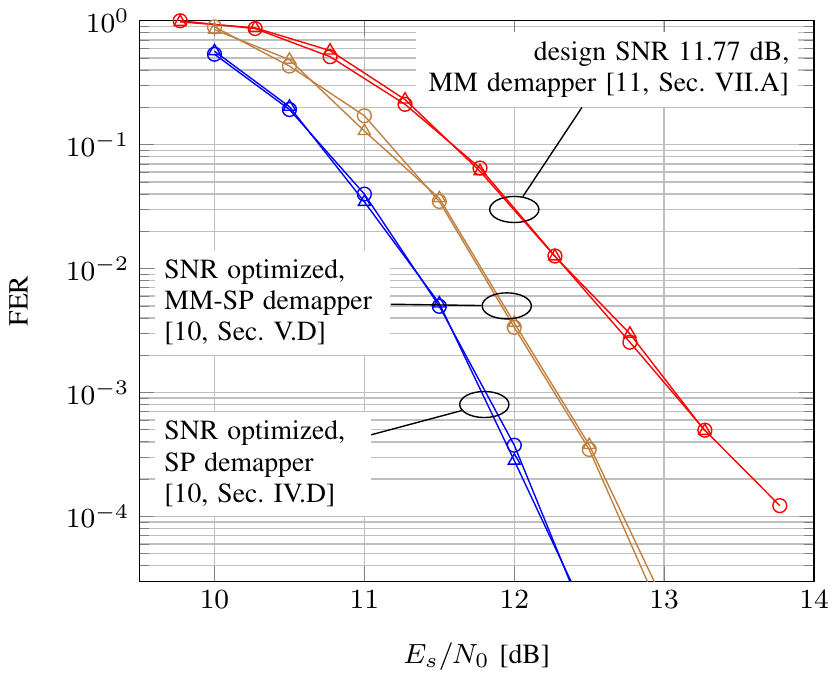}
\parbox{0.6\columnwidth}{
circle: MC construction.\\
triangle: LM-DGA construction.
}
\caption{Comparison of the proposed LM-DGA construction with the MC construction for several demappers from literature. Frame error rates (FER) under SC decoding are displayed. Rate 1/2 codes with block length 1536 bits are constructed for 64-QAM constellations. The resulting spectral efficiency is 3 bits/s/Hz.}
\label{fig:mainplot}
\end{figure}

In this work, we use the information-theoretic framework of mismatched decoding \cite{ganti2000mismatched,bocherer2016achievable}. We characterize the polar demapper bit channels by the LM-rate, which is an achievable rate under mismatched decoding. We then model the polar demapper bit channels by biAWGN channels with capacities equal to the LM-rates and use the GA construction. We evaluate this LM-DGA construction for several demappers proposed in literature. The LM-DGA constructed codes have the same performance as MC constructed codes, see Fig.~\ref{fig:mainplot}. For the MM-SP demapper \cite[Sec.~V.D]{seidl2013polar}, the proposed LM-DGA construction is about \SI{2}{dB} and \SI{1}{dB} more power efficient than the CGA and the MI-DGA construction, respectively, see Fig.~\ref{fig:constructions}. For 64-QAM, spectral efficiency 3 bits/s/Hz, and block length 1536 bits, simulation results show that LM-DGA constructed polar codes with cyclic redundancy check (CRC) outer codes and SC list decoding \cite{tal2015list} are \SI{1}{dB} more power efficient than state-of-the-art AR4JA \cite{divsalar2009capacity} low-density parity-check (LDPC) codes.

This work is organized as follows. In Sec.~\ref{sec:pcm}, we review PCM. We discuss achievable rates for polar demappers in Sec.~\ref{sec:airs}. In Sec.~\ref{sec:constructions}, we present the LM-DGA construction and compare it to the CGA and MI-DGA constructions. Sec.~\ref{sec:scl} provides numerical results for PCM with CRC and SC list decoding with comparison to LDPC codes. We conclude in Sec.~\ref{sec:conclusions}.

%%%%%%%%%%%%%%%%%%%
%%% POLAR CODED MODULATION
%%%%%%%%%%%%%%%%%%%

\section{Polar-Coded Modulation}
\label{sec:pcm}

\subsection{Channel Model}
We consider memoryless AWGN channels with bipolar amplitude shift keying (ASK) constellations and $2^m$ signal points given by
\begin{align}
\setx=\{\pm 1,\pm 3,\dotsc,\pm (2^m-1)\}.
\end{align}
The I/O relation of the AWGN channel is
\begin{align}
Y=X+\sigma Z\label{eq:awgn}
\end{align}
where $X$ is the channel input with distribution $P_X$ on $\setx$, $Y$ is the channel output and $Z$ is zero mean Gaussian noise with variance one. The SNR is $\expop[X^2]/\sigma^2$. Note that two real ASK symbols are equivalent to one complex quadrature amplitude modulation (QAM) symbol.

\subsection{Polar Coding}
\begin{figure}
\centering
\footnotesize
%\tikzsetnextfilename{size4PolarCodeFG}
%\input{figures/size4PolarCodeFG}
\includegraphics{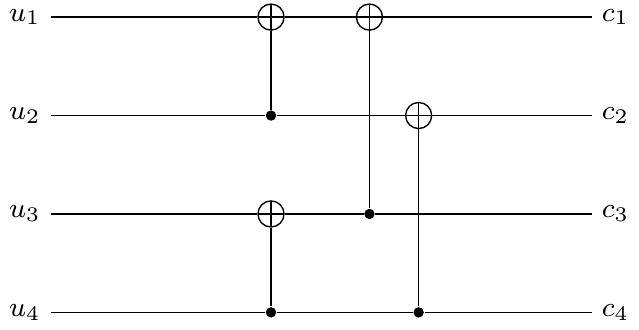}
\caption{Factor graph representation of the polar transformation $\mathbb{F}^{\logn}$ for $\nc=4$.}
\label{fig:fg}
\end{figure}
A binary polar code of block length $n$ and dimension $k$ is defined by $n-k$ frozen positions and the polar transformation $\mathbb{F}^{\otimes\logn}$, which denotes the $\logn$-fold Kronecker power of the transform
\begin{equation}
\mathbb{F}= \begin{bmatrix}
1 & 0\\
1 & 1
\end{bmatrix}.
\end{equation}
Polar encoding can be represented by
\begin{align}
\vecu\mathbb{F}^{\otimes\logn}=\vecc
\end{align}
where the $n-k$ frozen positions in $\vecu$ are set to predetermined values and where the unfrozen positions contain the $k$ information bits. The vector $\vecc$ is the code word. Factor graphs \cite{kschischang2001factor} are a convenient representation of polar codes, see Fig.~\ref{fig:fg} for $\nc=4$. SC decoding estimates the bits $u_1u_2\dotsc u_{\nc}$ successively, i.e., the channel output $y^n=y_1y_2\dotsc y_n$ and the estimates $\hat{u}_1\dotsc \hat{u}_i$ are used to estimate bit $u_{i+1}$. Encoding and decoding can be performed with $\mathcal{O}(\nc\log\nc)$ complexity \cite[Sec.~5.2.2]{stolte2002rekursive}.

\subsection{Polar Mapper and Demapper}
\begin{figure}
  \footnotesize
%\tikzsetnextfilename{pcm2}
%\input{figures/pcm2}
\includegraphics{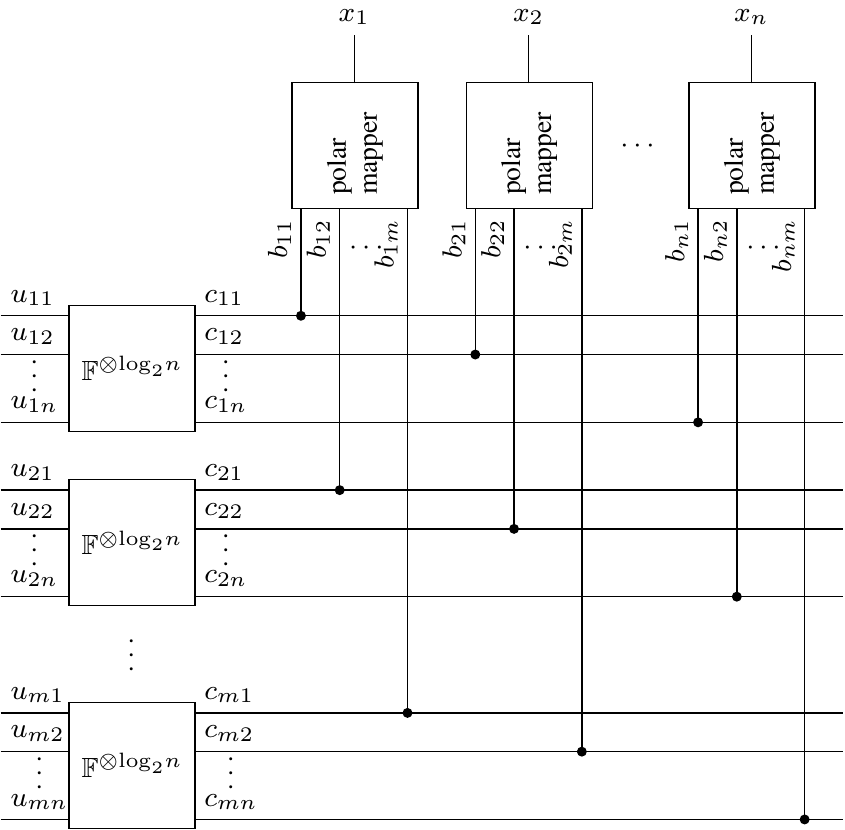}
\caption{Canonical PCM \cite{seidl2013polar}.}
\label{fig:pcm}
\end{figure}
Fig.~\ref{fig:pcm} displays the canonical PCM \cite{seidl2013polar} scheme for $2^m$-ASK constellations. Encoding works as follows. The length $m\nc$ vector $\vecu$ consisting of information bits and frozen bits is split into $m$ vectors $\vecu_1,\dotsc,\vecu_m$, which are then mapped to $m$ vectors $\vecc_j=\vecu_j\mathbb{F}^{\otimes\logn}$. A polar mapper implements a label function that maps the $m$ bits $c_{1i}\dotsc c_{mi}$ to the $i$th transmitted ASK symbol $x_i$ for $i=1,2,\dotsc,\nc$, i.e., for $j=1,\dotsc,m$, the output $\vecc_j$ of the $j$th polar transformation is mapped to the $j$th bit level of the labeling function.
\begin{figure}
  \footnotesize
  \centering
%\tikzsetnextfilename{pdem}
%\input{figures/pdem}
\includegraphics{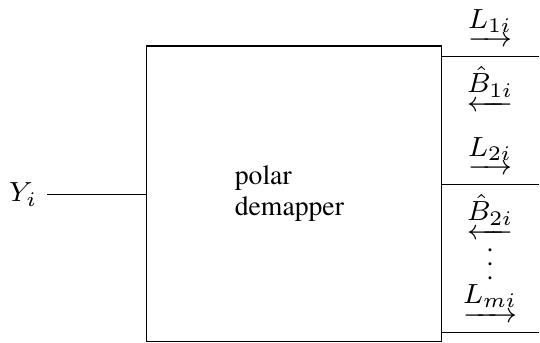}
  \caption{Generic polar demapper. See \eqref{eq:pdem} for a formal definition.}
\label{fig:pdem}
\end{figure}
We define the polar label
\begin{align}
\vecb_i=b_{i1}\dotsc b_{im}:=c_{1i}\dotsc c_{mi},\quad i=1,2,\dotsc,n.
\end{align}
In the following, we will for notational convenience sometimes drop the index $i$ and write $\vecb=b_1\dotsc b_m$ to refer to the polar label at a generic time instance.

Decoding works according to the schedule  
\begin{align}\label{eq:pcm}
\footnotesize
\begin{split}
&\text{1.1) demap polar level 1}\\
&\text{1.2) decode polar level 1}\\
&\text{2.1) demap polar level 2}\\
&\text{2.2) decode polar level 2}\\
&\vdots\\
&\text{$m$.1) demap polar level $m$}\\
&\text{$m$.2) decode polar level $m$}.
\end{split}
\end{align}
The polar demapper displayed in Fig.~\ref{fig:pdem} passes soft-information $L_{ji}$ to the $j$th polar decoder, which returns its estimate $\hat{B}_{ji}$. The polar demapper successively calculates
\begin{equation}\label{eq:pdem}
  \begin{split}
  L_{1i}&=\lambda_1(Y_i)\\
  L_{2i}&=\lambda_2(Y_i,\hat{B}_{1i})\\
    &\vdots\\
  L_{mi}&=\lambda_m(Y_i,\hat{B}_{1i},\dots,\hat{B}_{(m-1)i}).
\end{split}
\end{equation}
\begin{table}
\centering
\caption{Polar Mappers}
\label{tab:labels}
\begin{tabular}{rcccccccc}
\toprule
8-ASK symbols&-7&-5&-3&-1&1&3&5&7\\\midrule
&\multicolumn{8}{c}{MM Mapper \cite[Sec.~VII.A]{mahdavifar2016polar}}\\\midrule
BRGC $\tilde{\vecb}$&000&001&011&010&110&111&101&100\\
polar label $\vecb$&000&111&001&110&010&101&011&100\\
\midrule
&\multicolumn{8}{c}{SP Mapper}\\\midrule
LSB-BRGC $\tilde{\vecb}$&000&100&110&010&011&111&101&001\\
SP polar label $\vecb$&000&100&010&110&001&101&011&111\\\bottomrule
\end{tabular}
\vspace{0.5cm}
\caption{Label Transform of MM Mapper \cite[Sec.~VII.A]{mahdavifar2016polar} and SP Mapper}
\label{tab:label trafo}
\begin{tabular}{cm{3cm}m{3cm}}
&$\tilde{\vecb}=\vecb\bbm 1&0&0\\1&1&0\\0&1&1\ebm$&$\vecb=\tilde{\vecb}\bbm 1&0&0\\1&1&0\\1&1&1\ebm$\end{tabular}
\end{table}
\subsection{Polar Demappers for 8-ASK}
\begin{figure}
\footnotesize
%\tikzsetnextfilename{pdem_mahdavifar}
%\input{figures/pdem_mahdavifar}
\includegraphics{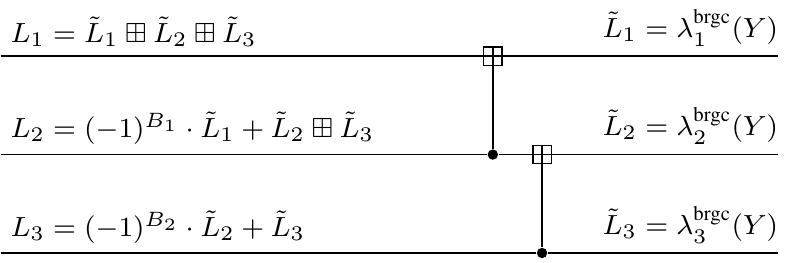}
  \caption{MM demapper proposed in \cite[Sec.~VII.A]{mahdavifar2016polar}. $\tilde{B}_1\tilde{B}_2\tilde{B}_3$ is a BRGC. The resulting polar label $B_1B_2B_3$ is displayed in Table~\ref{tab:labels}.}
  \label{fig:pdem_mahdavifar}

%\tikzsetnextfilename{pdem_mm-sp}
%\input{figures/pdem_mm-sp}
\includegraphics{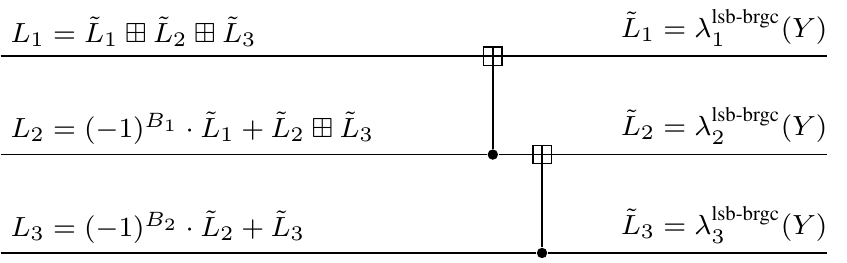}
  \caption{MM-SP demapper adapted from \cite[Sec.~V.D]{seidl2013polar}. $B_1B_2B_3$ is an SP polar label and $\tilde{B}_1\tilde{B}_2\tilde{B}_3$ is an LSB-BRGC, see Table~\ref{tab:labels}.}
  \label{fig:pdem_mm-sp}
\vspace{0.2cm}
\hrule

\begin{align*}
  &L_1=\log\frac{p_{B_1|Y}(0|Y)}{p_{B_1|Y}(1|Y)}\\
  &L_2=\log\frac{p_{B_2|YB_1}(0|YB_1)}{p_{B_2|YB_1}(1|YB_1)}\\
  &L_3=\log\frac{p_{B_3|YB_1B_2}(0|YB_1B_2)}{p_{B_3|YB_1B_2}(1|YB_1B_2)}
\end{align*}
\hrule
  \caption{SP demapper adapted from \cite[Sec.~IV.D]{seidl2013polar}. $B_1B_2B_3$ is an SP label, see Table~\ref{tab:labels}.}
  \label{fig:pdem_sp}
\end{figure}

We next present three polar demappers for 8-ASK that have been proposed in literature.
\subsubsection{MM Demapper \cite[Sec.~VII.A]{mahdavifar2016polar}}

The mapper proposed in \cite[Sec.~VII.A]{mahdavifar2016polar} is displayed in Table~\ref{tab:labels}. The polar label $\vecb$ is first mapped to the Binary Reflected Gray Code (BRGC)~\cite{gray1953pulse} $\tilde{\vecb}$, which is then mapped to an 8-ASK symbol. We show the label transformation in Table~\ref{tab:label trafo}. The MM demapper is displayed in Fig.~\ref{fig:pdem_mahdavifar}. The $\tilde{L}_j$ are calculated as
\begin{align}
\tilde{L}_j=\lambda_j^\text{brgc}(Y)=\log\frac{P_{\tilde{B}_j|Y}(0|Y)}{P_{\tilde{B}_j|Y}(1|Y)},\qquad j=1,2,3.
\end{align}
The $\tilde{L}_j$ are then combined according to Fig.~\ref{fig:pdem_mahdavifar}. The boxplus operation \cite[Eq.~(11)]{hagenauer1996iterative} is defined by
\begin{align}
L\boxplus L':= 2\atanh\left[\tanh\left(\frac{L}{2}\right)\tanh\left(\frac{L'}{2}\right)\right].
\end{align}
The $\tilde{L}_j$, $j=1,2,3$ are calculated from the same channel output $Y$ and are therefore stochastically dependent. This is ignored by the boxplus operation, which assumes independence. Consequently,
\begin{align}
L_j\neq\log\frac{P_{B_j|YB_1^{j-1}}(0|YB_1^{j-1})}{P_{B_j|YB_1^{j-1}}(1|YB_1^{j-1})}, \quad j=1,2,3
\end{align}
where $B_1^{j-1}=B_1\dotsc B_{j-1}$. The MM demapper is therefore \emph{mismatched}. We discuss achievable rates for mismatched decoding in Sec.~\ref{subsec:mismatch}.

\subsubsection{MM-SP Demapper \cite[Sec.~V.D]{seidl2013polar}} In \cite[Sec.~V.D]{seidl2013polar}, the set partitioning (SP) \cite{ungerbock1982channel} mapper is proposed for 16-ASK. The corresponding SP mapper for 8-ASK is displayed in Table~\ref{tab:labels}. A polar SP label is mapped to a least significant bit (LSB) BRGC, which is then mapped to an 8-ASK symbol. We show the label transformation in Table~\ref{tab:label trafo}. The demapper is shown in Fig.~\ref{fig:pdem_mm-sp}. This demapper is also \emph{mismatched}.

\subsubsection{SP Demapper \cite[Sec.~IV.D]{seidl2013polar}}
The SP demapper calculates the soft-information $L_j$ for the polar SP label according to Fig.~\ref{fig:pdem_sp}. The formulas in Fig.~\ref{fig:pdem_sp} directly imply that the SP demapper is \emph{matched}.
\begin{remark}
The SP demapper is equivalent to successively calculating the $L$-values of one 8-ASK bit level, one 4-ASK bit level, and one 2-ASK bit level. The two mismatched demappers first calculate the $L$-values of three 8-ASK bit levels and then calculate in addition the two boxplus operations $L_1\boxplus L_2$ and $(L_1\boxplus L_2)\boxplus L_3$. Consequently, the SP demapper has \emph{lower} complexity than the mismatched demappers.
\end{remark}

%%%%%%%%%%%%%%%%%%%
%%% ACHIEVABLE RATES
%%%%%%%%%%%%%%%%%%%

\section{Achievable Rates of Polar Demappers}
\label{sec:airs}

\subsection{Mutual Information}
Mutual information is an achievable rate for reliable communication \cite[Sec.~5.6]{gallager1968information}. Consider a memoryless channel $p_{Y|X}$ and a random codebook $\setc=\{X^n(1),\dotsc,X^n(2^{nR})\}$ with entries independent and identically distributed according to $P_X$ on $\setx$. The message $W$ is uniformly distributed on $\{1,2,\dotsc,2^{nR}\}$, i.e., the rate is $\log_2(2^{nR})/n=R$ bits per channel use. A \emph{maximum likelihood} (ML) decoder uses $p_{Y|X}(\cdot|\cdot)$ as decoding metric, i.e.,
\begin{align}
\hat{W}=\argmax_{w\in\{1,2,\dotsc,2^{nR}\}} \prod_{i=1}^n p_{Y|X}(Y_i|X_i(w)).
\end{align}
The average probability of erroneous ML decoding $\Pr(W\neq\hat{W})$ of the random code ensemble approaches zero for $n$ approaching infinity if
\begin{align}
R<\miop(X;Y).
\end{align}

\subsection{LM-Rate}
\label{subsec:mismatch}
Using a metric different from $p_{Y|X}$ is called \emph{mismatched decoding} \cite{ganti2000mismatched}. A mismatched decoder uses a function $q(\cdot,\cdot)$, called \emph{auxiliary channel}, as decoding metric, i.e.,
\begin{align}
\widehat{W}=\argmax_{w\in\{1,2,\dotsc,2^{nR}\}} \prod_{i=1}^n q(Y_i,X_i(w)).
\end{align}
Define
\begin{align}
\rate(X,Y,q,r,s)=\expop\left[\log_2\frac{q(Y,X)^sr(X)}{q_{r,s}(Y)}\right]\label{eq:lmrate}
\end{align}
where $q_{r,s}(\cdot)=\expop[q(\cdot,X)^s r(X)]$ is the corresponding auxiliary output distribution, where $s\geq 0$, and where $r\colon\setx\to\setr$ is a real-valued function defined on $\setx$ with finite expectation $\expop[r(X)]<\infty$. By \cite{ganti2000mismatched}, the probability of erroneous mismatched decoding $\Pr(W\neq\widehat{W})$ approaches zero for $n$ approaching infinity if
\begin{align}
R<\rate(X,Y,q,r,s).\label{eq:lmrate achievable}
\end{align}
\begin{lemma}
We have $\rate(X,Y,q,r,s)\leq\miop(X;Y)$, with equality if $q=p_{Y|X}$ and $s=1$, $r(\cdot)=1$.
\end{lemma}
Maximizing over $r,s$ yields the $\emph{LM-Rate}$ \cite{ganti2000mismatched}
\begin{align}
\rlm(X,Y,q):=\max_{r,s}\rate(X,Y,q,r,s).
\end{align}
\begin{remark}
Setting $r(\cdot)=1$ and maximizing over $s$ yields the \emph{generalized mutual information} (GMI) as defined in \cite{kaplan1993information}.
\end{remark}

\subsection{LM-Rate for Polar Demappers}

We now evaluate the LM-Rate for the metric
\begin{align}
\qbi(L,B):=e^{-\frac{L}{2}(1-2B)}\label{eq:metric LB}
\end{align}
where $L$ is the demapper output providing soft-information about bit $B$. For the metric \eqref{eq:metric LB}, the LM-rate becomes
\begin{align}
&\rlm(B,L,\qbi) \nonumber\\
&=\max_{r,s}\expop\left[\log_2\frac{e^{s\frac{L}{2}(1-2B)}r(B)}{\sum_{b\in\{0,1\}}P_B(b)e^{s\frac{L}{2}(1-2b)}r(b)}\right].
\end{align}
\begin{lemma}\label{lem:matched}
For
\begin{align}
 L = \log\frac{P_{B|Y}(0|Y)}{P_{B|Y}(1|Y)},\quad s=1,\quad r(b)=\frac{1}{P_B(b)}
\end{align}
we have
\begin{align}
\rlm(B,L,\qbi,r,s)=\miop(B;Y).
\end{align}
\end{lemma}
\begin{IEEEproof}
We provide a proof in the Appendix.
\end{IEEEproof}
We can now estimate an achievable rate for each bit level $j=1,2,\dotsc,m$ of a polar demapper. It is given by
\begin{align}
\rlm\Bigl[B_j,\lambda_j(Y,B_1^{j-1}),\qbi\Bigr].
\end{align}
\subsection{Polar Demapper Achievable Rates for 8-ASK}
In Table~\ref{tab:airs}, we display the LM-rates at \SI{11.77}{dB} for the bit-channels created by the MM demapper, the MM-SP demapper, and the SP demapper. The MM-SP and the SP demapper polarize more than the MM demapper. The SP demapper has the greatest sum rate. Note that this qualitatively corresponds to the ordering of the corresponding FER curves in Fig.~\ref{fig:mainplot}. For the MM demapper and the MM-SP demapper, we also display the MIs of the bit-channels before polar demapping. Both demappers have the same MIs in different order. Note that the bit-channels before polar demapping are much less polarized than after polar demapping.

\begin{table}
\caption{Polar Demapper Achievable Rates for \SI{11.77}{dB}}
\label{tab:airs}
\begin{tabular}{rcccc}
\toprule
\multicolumn{5}{c}{Polar demapper bit-channel achievable rates $\rlm(B_j,L_j,\qbi)$}\\
&$R_1$&$R_2$&$R_3$&$\sum_j R_j$\\\midrule
MM Demapper Fig.~\ref{fig:pdem_mahdavifar}&0.1294&0.9109&0.8212&1.8615\\
MM-SP Demapper Fig.~\ref{fig:pdem_mm-sp}&0.1295&0.7397&0.9885&1.8577\\
SP Demapper Fig.~\ref{fig:pdem_sp}&0.1312&0.7503&0.9986&1.8801\\
\midrule
&\multicolumn{4}{c}{Bit-channel achievable rates $\miop(\tilde{B}_j,\tilde{L}_j)$}\\
&$\tilde{R}_1$&$\tilde{R}_2$&$\tilde{R}_3$&$\sum_j \tilde{R}_j$\\\midrule
MM Demapper Fig.~\ref{fig:pdem_mahdavifar}&0.8319&0.6641&0.3589&1.8548\\
MM-SP Demapper Fig.~\ref{fig:pdem_mm-sp}&0.3589&0.6641&0.8319&1.8548\\\bottomrule
\end{tabular}
\end{table}

\section{Construction by Surrogates}
\label{sec:constructions}

\subsection{biAWGN Surrogate Channel}

The biAWGN channel is \eqref{eq:awgn} for $m=1$, i.e.,
\begin{align}
Y=x_b+\sigma Z
\end{align}
where $x_0=1$ and $x_1=-1$. The mutual information of input $B$ uniformly distributed on $\{0,1\}$ and biAWGN output $Y$ is
\begin{align}
\rbiawgn(\sigma^2)&=\miop(B;x_B+\sigma Z).
\end{align}

\subsection{Gaussian Approximation}

\begin{figure}
\footnotesize
\centering
%\tikzsetnextfilename{size2PolarCode}
%\include{figures/size2PolarCode}
\includegraphics{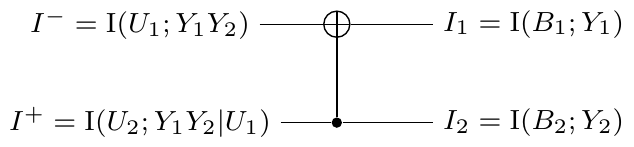}
\caption{MIs of the basic polar transform.}
\label{fig:PolarCodeCheckVariableNode}
\end{figure}

The reliability of the bit $U_i$, $i=1,2,\dots,n$ can be quantified by the MI $\miop(U_i;Y^n|U_1^{i-1})$. We can calculate these MIs by recursively calculating the MIs of the basic polar transform displayed in Fig.~\ref{fig:PolarCodeCheckVariableNode}. 
For a biAWGN channel, the update rule for the basic polar transform is given by \cite{tenbrink2004design}
\begin{align}
I^- &= 1- J\left(\sqrt{\left[J^{-1} (1-I_1)\right]^2 + \left[J^{-1} (1-I_2)\right]^2}\right) \label{eq:GA1} \\
I^+ &= J\left(\sqrt{\left[J^{-1} (I_1)\right]^2 + \left[J^{-1} (I_2)\right]^2}\right) \label{eq:GA2}
\end{align}
where the $J$-function is
\begin{equation} \label{eq:JFunction}
J(\sigma) = 1- \int_{-\infty}^{+\infty} \frac{e^{-\left(\left(\xi - \sigma^2/2\right)^2 /2\sigma^2\right)}}{\sqrt{2\pi}\sigma} \cdot \log_2 \left(1 + e^{-\xi}\right) \mathrm{d}\xi.
\end{equation}
To calculate \eqref{eq:JFunction} and its inverse, we use
\begin{equation}
J(\sigma) \approx \left(1-2^{-H_1\sigma^{2H_2}}\right)^{H_3} 
\end{equation}
and \begin{equation}
J^{-1}(I) \approx \left(-\frac{1}{H_1}\log_2 \left(1-I^{\frac{1}{H_3}}\right)\right)^{\frac{1}{2H_2}}
\end{equation}
from \cite[Eqs. (9),(10)]{brannstrom2005convergence} where $H_1 = 0.3073$, $H_2 = 0.8935$ and $H_3 = 1.1064$.

\subsection{Construction Methods}

\begin{figure}
\centering
\footnotesize
%\tikzsetnextfilename{surrogates}
%\input{figures/surrogates}
\includegraphics{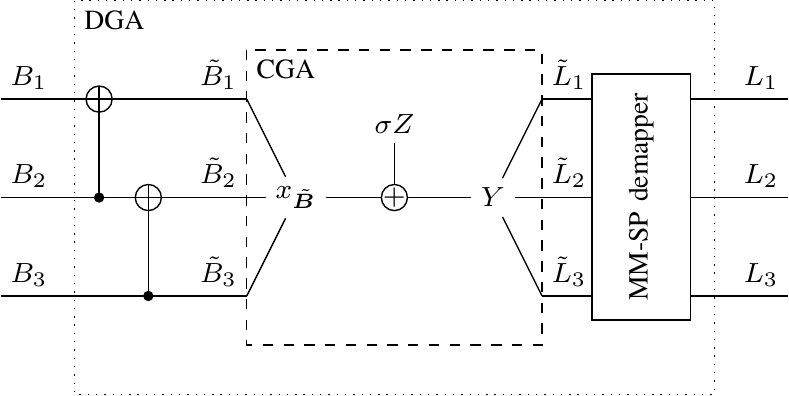}
\caption{Illustration of CGA and DGA for the MM-SP demapper from Fig.~\ref{fig:pdem_mm-sp}. CGA (DGA) replaces the inputs and outputs of the dashed (dotted) box by biAWGN channels.}
\label{fig:surrogates}
\end{figure}
We next discuss three variants of the GA construction. For illustrative purpose, we explain it for the 8-ASK MM-SP demapper. In Fig.~\ref{fig:surrogates}, MM-SP mapper, AWGN channel, and MM-SP demapper are displayed. The CGA construction connects $\tilde{B}_j$ and $\tilde{L}_j$ by a biAWGN channel with noise variance $\sigma_j^2$ given by
\begin{align}
\sigma^2_j\colon \rbiawgn(\sigma^2_j)=\miop(\tilde{B}_j;\tilde{L}_j).
\end{align}
The MI-DGA construction connects $B_j$ and $L_j$ by a biAWGN channel with noise variance
\begin{align}
\sigma^2_j\colon \rbiawgn(\sigma^2_j)=\miop(B_j;L_j).
\end{align}
The LM-DGA construction connects $B_j$ and $L_j$ by a biAWGN channel with noise variance
\begin{align}
\sigma^2_j\colon \rbiawgn(\sigma^2_j)=\rlm(B_j,L_j,\qbi).
\end{align}
\begin{figure}
\footnotesize
%\tikzsetnextfilename{constructions}
%\input{figures/constructions}
\includegraphics{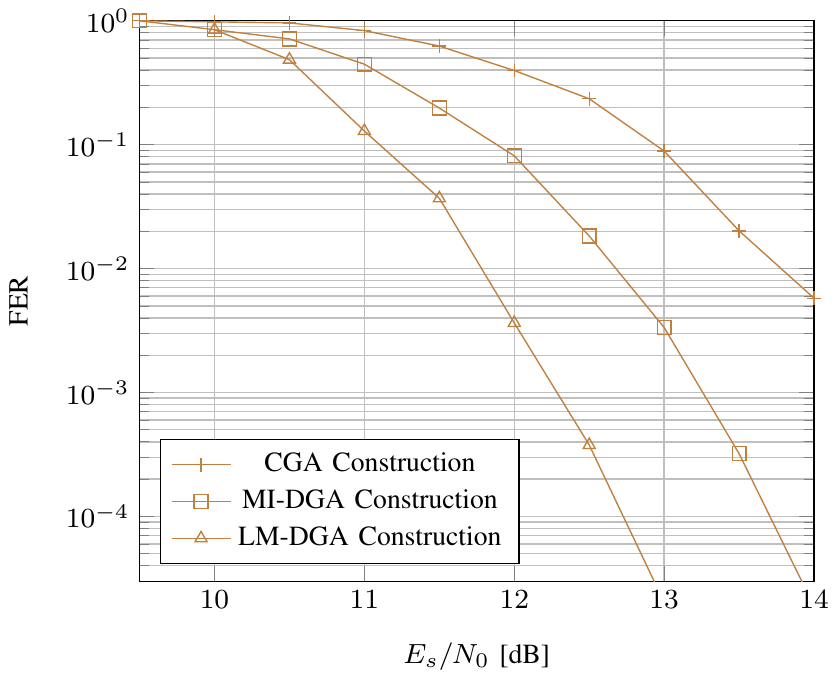}
\caption{Comparison of the CGA, MI-DGA, and LM-DGA constructions for the MM-SP demapper. Frame error rates (FER) under SC decoding are displayed. Rate 1/2 codes with block length 1536~bits are constructed for 64-QAM constellations. The resulting spectral efficiency is 3 bits/s/Hz.}
\label{fig:constructions}
\end{figure}
In Fig.~\ref{fig:constructions}, we show the FER performance for the three construction methods for the MM-SP demapper. The LM-DGA constructed code performs best and the CGA construction performs worst.

\subsection{SC List Decoding}
\label{sec:scl}
\begin{figure}
\centering
\footnotesize
%\tikzsetnextfilename{scl}
%\input{figures/scl}
\includegraphics{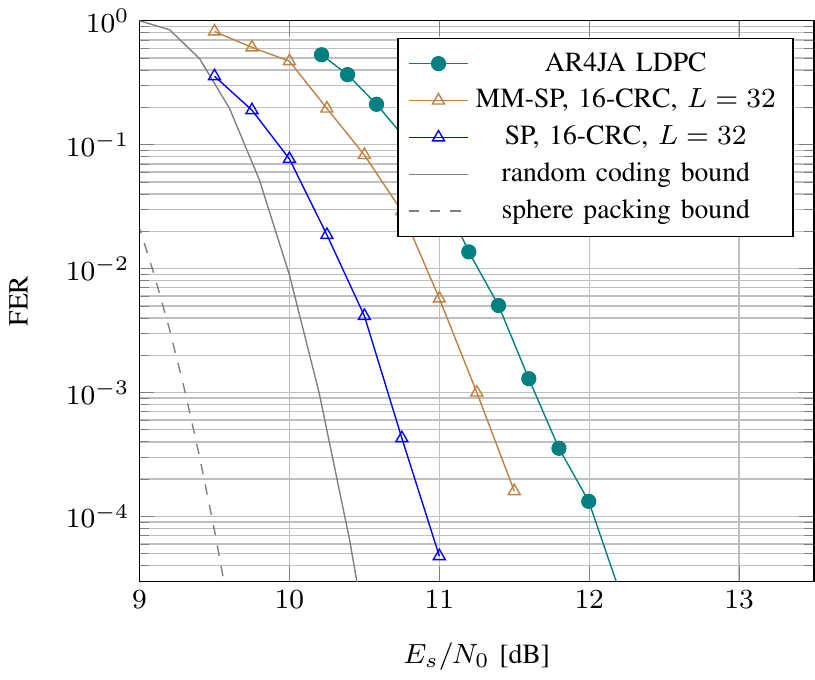}
\caption{LM-DGA constructed polar codes are combined with 16-CRC and SC list decoding with list size $L=32$ is applied. The SP demapper performs best and is around \SI{1}{dB} more power efficient than an AR4JA \cite{divsalar2009capacity} LDPC code decoded with 200 iterations. Rate 1/2 codes with block length 1536~bits are used for 64-QAM constellations. The resulting spectral efficiency is 3~bits/s/Hz.}
\label{fig:scl}
\end{figure}

In Fig.~\ref{fig:scl}, we show results for LM-DGA constructed polar codes combined with 16-CRC and SC list decoding \cite{tal2015list} with list size $L=32$. As a reference, Shannon's sphere packing bound~\cite[Eqs. (3), (17)]{shannon_probability_1959} and Gallager's random coding bound~\cite[Theorem~5.6.2]{gallager1968information} are shown as well. Both with an MM-SP and an SP demapper, the polar codes perform better than an AR4JA LDPC code \cite{divsalar2009capacity} decoded with 200 full sum-product belief propagation iterations. At FER $=10^{-4}$, the polar code with SP demapper is \SI{1}{dB} more power efficient than the LDPC code and is within \SI{0.5}{dB} of the random coding bound.

%%%%%%%%%%%%%%%%%%%
%%% CONCLUSIONS
%%%%%%%%%%%%%%%%%%%

\section{Conclusions}
\label{sec:conclusions}

In this work, we have developed the LM-rate demapper Gaussian approximation (LM-DGA) method to construct polar codes for higher-order modulation. We have shown that in contrast to GA construction methods previously proposed in literature, the LM-DGA construction works also for mismatched demappers, i.e., the LM-DGA constructed polar codes have the same performance as codes constructed by Monte Carlo simulation. With CRC outer codes and list decoding, the LM-DGA constructed polar codes outperform state-of-the-art LDPC codes. An interesting problem for future research are performance guarantees for mismatched demappers.

%%%%%%%%%%%%%%%%%%%
%%% APPENDIX
%%%%%%%%%%%%%%%%%%%

\appendix

\begin{IEEEproof}[Proof of Lemma~\ref{lem:matched}] For
\begin{align}
L = \log\frac{P_{B|Y}(0|Y)}{P_{B|Y}(1|Y)},\quad s=1,\quad r(b)=\frac{1}{P_B(b)}
\end{align}
we have
\begin{align}
P_{B|Y}(0|Y)=\frac{e^{L}}{1+e^L},\quad P_{B|Y}(1|Y)=\frac{1}{1+e^L}.
\end{align}
We have
\begin{align}
&\rlm(B,L,\qbi,r,s)\nonumber\\
&\qquad=\expop\left[\log_2\frac{e^{\frac{L}{2}(1-2B)}\frac{1}{P_B(B)}}{\sum_{b\in\{0,1\}}P_B(b)e^{\frac{L}{2}(1-2b)}\frac{1}{P_B(b)}}\right]\\
&\qquad=\entop(B)+\expop\left[\log_2\frac{e^{\frac{L}{2}(1-2B)}}{\sum_{b\in\{0,1\}}e^{\frac{L}{2}(1-2b)}}\right].
\end{align}
Continuing the last equation, we have
\begin{align}
&=\entop(B)+\expop\left[\log_2\frac{e^{\frac{L}{2}(1-2B)}}{e^{\frac{L}{2}}+e^{-\frac{L}{2}}}\right]\\
&=\entop(B)-\entop(B|Y)=\miop(B;Y).
\end{align}
\end{IEEEproof}
\bibliographystyle{IEEEtran}
\bibliography{IEEEabrv,confs-jrnls-DONTEDIT,references-DONTEDIT,references}

% Generated by IEEEtran.bst, version: 1.13 (2008/09/30)
\begin{thebibliography}{10}
\providecommand{\url}[1]{#1}
\csname url@samestyle\endcsname
\providecommand{\newblock}{\relax}
\providecommand{\bibinfo}[2]{#2}
\providecommand{\BIBentrySTDinterwordspacing}{\spaceskip=0pt\relax}
\providecommand{\BIBentryALTinterwordstretchfactor}{4}
\providecommand{\BIBentryALTinterwordspacing}{\spaceskip=\fontdimen2\font plus
\BIBentryALTinterwordstretchfactor\fontdimen3\font minus
  \fontdimen4\font\relax}
\providecommand{\BIBforeignlanguage}[2]{{%
\expandafter\ifx\csname l@#1\endcsname\relax
\typeout{** WARNING: IEEEtran.bst: No hyphenation pattern has been}%
\typeout{** loaded for the language `#1'. Using the pattern for}%
\typeout{** the default language instead.}%
\else
\language=\csname l@#1\endcsname
\fi
#2}}
\providecommand{\BIBdecl}{\relax}
\BIBdecl

\bibitem{stolte2002rekursive}
N.~Stolte, ``{Rekursive Codes mit der Plotkin-Konstruktion und ihre
  Decodierung},'' Ph.D. dissertation, TU Darmstadt, 2002.

\bibitem{arikan2009channel}
E.~Ar{\i}kan, ``Channel polarization: A method for constructing
  capacity-achieving codes for symmetric binary-input memoryless channels,''
  \emph{{IEEE} Trans. Inf. Theory}, vol.~55, no.~7, pp. 3051--3073, Jul. 2009.

\bibitem{vangala2015comparative}
\BIBentryALTinterwordspacing
H.~Vangala, E.~Viterbo, and Y.~Hong, ``A comparative study of polar code
  constructions for the {AWGN} channel,'' \emph{arXiv preprint}, 2015.
  [Online]. Available: \url{https://arxiv.org/pdf/1501.02473v1.pdf}
\BIBentrySTDinterwordspacing

\bibitem{mori2009performance}
R.~Mori and T.~Tanaka, ``Performance of polar codes with the construction using
  density evolution,'' \emph{{IEEE} Commun. Lett.}, vol.~13, no.~7, pp.
  519--521, Jul. 2009.

\bibitem{trifonov2012efficient}
P.~Trifonov, ``Efficient design and decoding of polar codes,'' \emph{{IEEE}
  Trans. Commun.}, vol.~60, no.~11, pp. 3221--3227, Nov. 2012.

\bibitem{dosio2016polar}
D.~Dosio, ``Polar codes for error correction: analysis and decoding
  algorithms,'' Master's thesis, University of Bologna, 2016.

\bibitem{chung2001analysis}
S.~Y. Chung, T.~J. Richardson, and R.~L. Urbanke, ``Analysis of sum-product
  decoding of low-density parity-check codes using a {Gaussian}
  approximation,'' \emph{{IEEE} Trans. Inf. Theory}, vol.~47, no.~2, pp.
  657--670, Feb. 2001.

\bibitem{tenbrink2004design}
S.~ten Brink, G.~Kramer, and A.~Ashikhmin, ``Design of low-density parity-check
  codes for modulation and detection,'' \emph{{IEEE} Trans. Commun.}, vol.~52,
  no.~4, pp. 670--678, Apr. 2004.

\bibitem{brannstrom2005convergence}
F.~Br\"annstr\"om, L.~K. Rasmussen, and A.~J. Grant, ``Convergence analysis and
  optimal scheduling for multiple concatenated codes,'' \emph{{IEEE} Trans.
  Inf. Theory}, vol.~51, no.~9, pp. 3354--3364, Sep. 2005.

\bibitem{seidl2013polar}
M.~Seidl, A.~Schenk, C.~Stierstorfer, and J.~B. Huber, ``Polar-coded
  modulation,'' \emph{{IEEE} Trans. Commun.}, vol.~61, no.~10, pp. 4108--4119,
  Oct. 2013.

\bibitem{mahdavifar2016polar}
H.~Mahdavifar, M.~El-Khamy, J.~Lee, and I.~Kang, ``Polar coding for
  bit-interleaved coded modulation,'' \emph{{IEEE} Trans. Veh. Technol.},
  vol.~65, no.~5, pp. 3115--3127, May 2016.

\bibitem{steiner2016protograph}
F.~Steiner, G.~B\"ocherer, and G.~Liva, ``Protograph-based {LDPC} code design
  for shaped bit-metric decoding,'' \emph{{IEEE} J. Sel. Areas Commun.},
  vol.~34, no.~2, pp. 397--407, Feb. 2016.

\bibitem{tavildar2016bit}
\BIBentryALTinterwordspacing
S.~R. Tavildar, ``Bit-permuted coded modulation for polar codes,'' \emph{arXiv
  preprint}, 2016. [Online]. Available:
  \url{https://arxiv.org/pdf/1609.09786v1.pdf}
\BIBentrySTDinterwordspacing

\bibitem{ganti2000mismatched}
A.~Ganti, A.~Lapidoth, and E.~Telatar, ``Mismatched decoding revisited: General
  alphabets, channels with memory, and the wide-band limit,'' \emph{{IEEE}
  Trans. Inf. Theory}, vol.~46, no.~7, pp. 2315--2328, Nov. 2000.

\bibitem{bocherer2016achievable}
\BIBentryALTinterwordspacing
G.~B\"ocherer, ``Achievable rates for shaped bit-metric decoding,'' \emph{arXiv
  preprint}, 2016. [Online]. Available: \url{http://arxiv.org/abs/1410.8075}
\BIBentrySTDinterwordspacing

\bibitem{tal2015list}
I.~Tal and A.~Vardy, ``List decoding of polar codes,'' \emph{{IEEE} Trans. Inf.
  Theory}, vol.~61, no.~5, pp. 2213--2226, May 2015.

\bibitem{divsalar2009capacity}
D.~Divsalar, S.~Dolinar, C.~R. Jones, and K.~Andrews, ``Capacity-approaching
  protograph codes,'' \emph{{IEEE} J. Sel. Areas Commun.}, vol.~27, no.~6, pp.
  876--888, Aug. 2009.

\bibitem{kschischang2001factor}
F.~R. Kschischang, B.~J. Frey, and H.-A. Loeliger, ``Factor graphs and the
  sum-product algorithm,'' \emph{{IEEE} Trans. Inf. Theory}, vol.~47, no.~2,
  pp. 498--519, Feb. 2001.

\bibitem{gray1953pulse}
F.~Gray, ``Pulse code communication,'' U. S. Patent 2\,632\,058, 1953.

\bibitem{hagenauer1996iterative}
J.~Hagenauer, E.~Offer, and L.~Papke, ``Iterative decoding of binary block and
  convolutional codes,'' \emph{{IEEE} Trans. Inf. Theory}, vol.~42, no.~2, pp.
  429--445, Mar. 1996.

\bibitem{ungerbock1982channel}
G.~Ungerb\"ock, ``Channel coding with multilevel/phase signals,'' \emph{{IEEE}
  Trans. Inf. Theory}, vol.~28, no.~1, pp. 55--67, Jan. 1982.

\bibitem{gallager1968information}
R.~G. Gallager, \emph{Information Theory and Reliable Communication}.\hskip 1em
  plus 0.5em minus 0.4em\relax John Wiley \& Sons, Inc., 1968.

\bibitem{kaplan1993information}
G.~Kaplan and S.~Shamai~(Shitz), ``Information rates and error exponents of
  compound channels with application to antipodal signaling in a fading
  environment,'' \emph{AE\"U}, vol.~47, no.~4, pp. 228--239, 1993.

\bibitem{shannon_probability_1959}
C.~E. Shannon, ``Probability of error for optimal codes in a {{Gaussian}}
  channel,'' \emph{Bell Syst. Tech.~J.}, vol.~38, no.~3, pp. 611--656, May
  1959.

\end{thebibliography}
\end{document}